# BCI-Controlled Hands-Free Wheelchair Navigation with Obstacle Avoidance*

Ramy Mounir, Redwan Alqasemi, and Rajiv Dubey, *Member, IEEE*

*Abstract*— Brain-Computer interfaces (BCI) are widely used in reading brain signals and converting them into real-world motion. However, the signals produced from the BCI are noisy and hard to analyze. This paper looks specifically towards combining the BCI's latest technology with ultrasonic sensors to provide a hands-free wheelchair that can efficiently navigate through crowded environments. This combination provides safety and obstacle avoidance features necessary for the BCI Navigation system to gain more confidence and operate the wheelchair at a relatively higher velocity. A population of six human subjects tested the BCI-controller and obstacle avoidance features. Subjects were able to mentally control the destination of the wheelchair, by moving the target from the starting position to a predefined position, in an average of 287.12 seconds and a standard deviation of 48.63 seconds after 10 minutes of training. The wheelchair successfully avoided all obstacles placed by the subjects during the test.

## I. INTRODUCTION

A Brain-Computer Interface (BCI) system allows the users to trigger various actions and communicate emotions without movement [2]. The most basic non-invasive BCI device consists of sensors (used to read electroencephalogram, EEG, signals), signal Amplifier, and a user interface. After the EEG signal has been digitized, it can be processed through various algorithms (P-300, SSVEP, ERD, Motor-Imagery, etc.) to extract the action/event used for control [2]. The choice of algorithm depends on the specific application, and in some cases, the medical condition of the user.

BCI systems have been used for numerous different applications, including brain painting [3], Quadcopter control [4], enhancing engagement levels [5] and controlling a 9-DoF wheelchair-mounted robotic arm [6]. BCI devices have also been used to control and navigate wheelchairs using only EEG signals [7]; however, safety features had to be added to avoid collision with obstacles. Additionally, mental fatigue resulting from the increased time-on-task (TOT) can "significantly affect the distribution of band power features" [8].

The BCI system implemented in this paper utilizes potential field algorithm to autonomously navigate to the destination position while avoiding obstacles. This implementation reduces the TOT and the mental fatigue associated with it. The system also requires the training of only one mental action to move the target around in a 2-D virtual space. Training one mental action significantly simplifies the training process and reduces the time required for training.

## II. SYSTEM DESIGN

An Invacare TDX Power Wheelchair was used as a prototype for this system. Grayhill 63R wheel Encoders were added to both wheels, and two sensor boxes (one on each side) were added to the wheelchair. A tablet was used to collect the data from the sensors, encoders and BCI device. The tablet is also used for all the processing, training of the BCI and displaying the Graphical User Interface (GUI). Output commands are sent from the tablet to the joystick module to override the manual inductive joystick and navigate the wheelchair.

### A. Brain-Computer Interface

This project uses active BCI (Motor Imagery) to provide a method for controlling the spatial position of a target destination for the wheelchair. Active BCI has been proven to result in a "higher overall task performance" – for wheelchair navigation - when compared to P300 reactive BCI conventional methods [7]. According to Stamps and Hamam [9], Emotiv Epoc is the most usable EEG acquisition device.

### B. Mobility Assist

This project uses eight ultrasonic sensors (four on each side) to detect obstacles. The moving average of three consecutive readings is calculated to reduce the noise. The positions of the obstacles relative to the wheelchair are used within the potential field algorithm to generation a repulsive force. However, the target position generates a larger attraction force. The resultant net force controls the translation and rotation components of the wheelchair navigation through a PID controller, while the encoders close the control loop. More information can be found in [1].

### C. Graphical User Interface

The user interface design allows the user to move the wheelchair manually using a virtual joystick or activate the auto drive mode which autonomously moves the wheelchair towards the target while avoiding obstacles. Other features include a visualization of the mental action power level and a level switch that allows the user to only activate the movement of the target beyond a specific brain power level. The GUI allows users to save a training profile and load it later to continue the training process.

The top virtual camera captures the main view of the GUI; however, a front view camera is also shown to help the users visualize the navigation process. The GUI also shows a vector representing the direction of forces (in Realtime)

*Research supported by Florida Department of Education, Division of - Vocational Rehabilitation.

Ramy Mounir is a PhD Candidate at the Department of Mechanical Engineering, University of South Florida, Tampa, FL 33620, USA (phone: (813) 397-9373, e-mail: ramy@mail.usf.edu).

Redwan Alqasemi, and Rajiv Dubey are faculty in the Department of Mechanical Engineering, University of South Florida, Tampa, FL 33620.

based on the potential field algorithm. Obstacles' positions are shown as red spheres while the target is represented with a green sphere. Figure 1 shows the GUI described above.

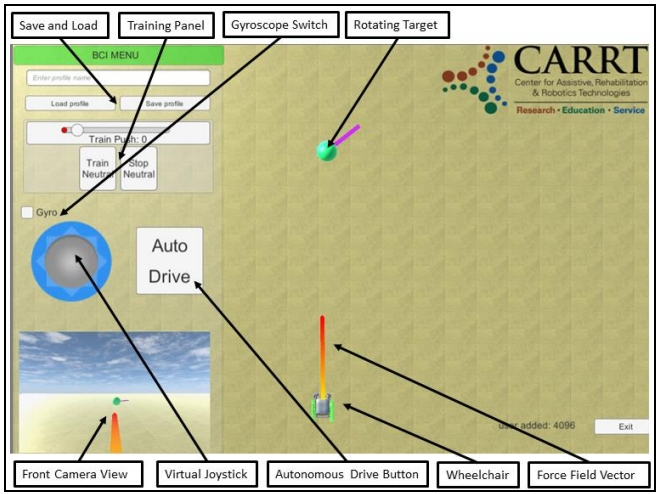

FIGURE 1: LABELED GRAPHICAL USER INTERFACE

## III. EVALUATIONS AND DISCUSSIONS

This project can be divided into 2 independent tasks; BCI and obstacle avoidance tasks. The BCI task is to move the target around on the GUI using only EEG signals, while the obstacle avoidance task is to autonomously move the wheelchair in a cluttered environment without colliding with obstacles. Six human subjects have tested the BCI and obstacle avoidance tasks.

Subjects were asked to train the BCI device for 10 minutes and move the target to a predefined destination. The target navigation times were recorded and plotted in figure 2, and the trails of the paths are shown in figure 3.

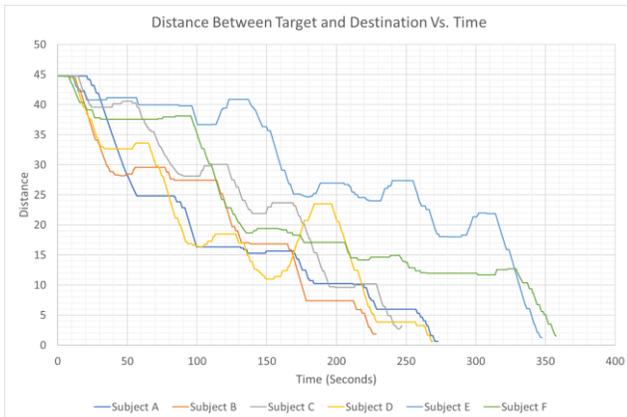

FIGURE 2: TARGET TO DESTINATION DISTANCE VS. TIME

Figure 2 shows that the subjects were able to successfully navigate the target to the destination. The distance continued to decrease with time, zero distance means that the target has reached its destination. Figure 3 shows the path of each target as controlled by the human subjects.

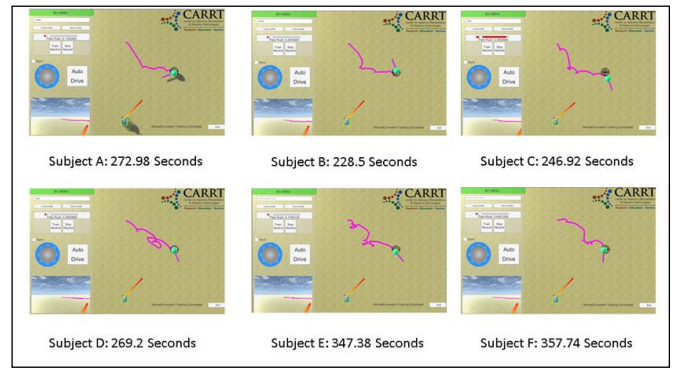

FIGURE 3: VISUALIZATION OF THE PATH CONTROLLED BY HUMAN SUBJECTS

## IV. CONCLUSION

The presented system offers full navigation control of the wheelchair around obstacles using the BCI. The PID control loop adjusts the translational speed of the wheelchair when close to an obstacle to ensure safety of the users. Users were able to successfully navigate the wheelchair completely through mental commands after only 10 minutes of motor imagery training. For better results, we plan on increasing the training time and implementing a more sophisticated obstacle avoidance algorithm.

REFERENCES

[1] Ashley, D., Ashley, K., Alqasemi, R., & Dubey, R. (2017, July). Semi-autonomous mobility assistance for power wheelchair users navigating crowded environments. In Rehabilitation Robotics (ICORR), 2017 International Conference on (pp. 1025-1030). IEEE.
[2] Wolpaw, J. R., Birbaumer, N., McFarland, D. J., Pfurtscheller, G., & Vaughan, T. M. (2002). Brain–computer interfaces for communication and control. Clinical neurophysiology, 113(6), 767-791.
[3] Botrel, L., Holz, E. M., & Kübler, A. (2015). Brain Painting V2: evaluation of P300-based brain-computer interface for creative expression by an end-user following the user-centered design. Brain-Computer Interfaces, 2(2-3), 135-149.
[4] LaFleur, K., Cassady, K., Doud, A., Shades, K., Rogin, E., & He, B. (2013). Quadcopter control in three-dimensional space using a noninvasive motor imagery-based brain–computer interface. Journal of neural engineering, 10(4), 046003.
[5] Andujar, M., & Gilbert, J. E. (2013, April). Let's learn!: enhancing user's engagement levels through passive brain-computer interfaces. In CHI'13 Extended Abstracts on Human Factors in Computing Systems (pp. 703-708). ACM.
[6] Palankar, M., De Laurentis, K. J., Alqasemi, R., Veras, E., Dubey, R., Arbel, Y., & Donchin, E. (2009, February). Control of a 9-DoF wheelchair-mounted robotic arm system using a P300 brain computer interface: Initial experiments. In Robotics and Biomimetics, 2008. ROBIO 2008. IEEE International Conference on (pp. 348-353). IEEE.
[7] Carlson, T., & Millan, J. D. R. (2013). Brain-controlled wheelchairs: a robotic architecture. IEEE Robotics & Automation Magazine, 20(1), 65-73.
[8] Roy, R. N., Bonnet, S., Charbonnier, S., & Campagne, A. (2013, July). Mental fatigue and working memory load estimation: interaction and implications for EEG-based passive BCI. In Engineering in Medicine and Biology Society (EMBC), 2013 35th Annual International Conference of the IEEE (pp. 6607-6610). IEEE.
[9] K. Stamps and Y. Hamam, "Towards Inexpensive BCI Control for Wheelchair Navigation in the Enabled Environment – A Hardware Survey," Brain Informatics Lecture Notes in Computer Science, pp. 336–345, 2010.
[10] BCI-Controlled Wheelchair : Obstacle Avoidance Demo, 30-Jul-2018. [Online]. Available: https://www.youtube.com/watch?v=cZN2oBABRjQ. [Accessed: 31-Jul-2018].